\documentclass{article}

\usepackage[english]{babel}
\usepackage[letterpaper,top=2cm,bottom=2cm,left=3cm,right=3cm,marginparwidth=1.75cm]{geometry}

\usepackage{amsmath}
\usepackage{graphicx}
\usepackage{subcaption}
\usepackage[colorlinks=true, allcolors=blue]{hyperref}
\usepackage{multirow} 
\usepackage{authblk}

\title{NCCR: to Evaluate the Robustness of Neural Networks and Adversarial Examples}
\author[1]{Shi Pu}
\author[2]{Fu Song}
\author[3]{Wenjie Wang\thanks{Corresponding author}}

\affil[1]{ShanghaiTech University \\ \texttt{pushi@shanghaitech.edu.cn}}
\affil[2]{State Key Laboratory of Computer Science, Institute of Software, Chinese Academy of Sciences \\ \texttt{songfu@ios.ac.cn}}
\affil[3]{ShanghaiTech University \\ \texttt{wangwj1@shanghaitech.edu.cn}}

\begin{document}
\maketitle

\begin{abstract}

Neural networks have received a lot of attention recently, and related security issues have come with it. Many studies have shown that neural networks are vulnerable to adversarial examples that have been artificially perturbed with modification, which is too small to be distinguishable by human perception. Different attacks and defenses have been proposed to solve these problems, but there is little research on evaluating the robustness of neural networks and their inputs. In this work, we propose a metric called the neuron cover change rate (NCCR) to measure the ability of deep learning models to resist attacks and the stability of adversarial examples. NCCR monitors alterations in the output of specifically chosen neurons when the input is perturbed, and networks with a smaller degree of variation are considered to be more robust. The results of the experiment on image recognition and the speaker recognition model show that our metrics can provide a good assessment of the robustness of neural networks or their inputs. It can also be used to detect whether an input is adversarial or not, as adversarial examples are always less robust.
\end{abstract}

\section{Introduction}

Deep learning(DL) has achieved  great success in many fields such as image recognition\cite{he2016deep}, natural language processing\cite{mikolov2013distributed} and speaker recognition\cite{kinnunen2010overview}. In addition to those success, the issue of limited robustness in neural network models is increasingly garnering attention. Many research surfaces that DL models are vulnerable to input with crafted small perturbations\cite{goodfellow2014explaining} which is called \emph{adversarial examples}. This is unacceptable when models are actually put into use especially in security-critical applications like autonomous driving\cite{kong2020physgan}, medical diagnostics\cite{shen2017deep} and malware detection\cite{yuan2014droid} systems.

Since the inception of the L-BFGS attack\cite{szegedy2013intriguing}, numerous adversarial attack methods have been proposed to explore the stability of DL models. Although many defense measures follow, most of them are breached by carefully designed adaptive attacks\cite{he2017adversarial} \cite{carlini2017towards}. Adversarial training works well for the attacks it is trained on, but it mostly fails for unseen attacks\cite{moosavi2017universal}. Modifying the model to defend the network, defensive distillation\cite{papernot2016distillation} for example,  enhanced the robustness of the network, making it more resilient to small perturbations in the image, but it was quickly defeated by newly proposed attack\cite{carlini2017towards}. There are also some efforts to distinguish adversarial examples from clean examples before feeding them into the network to achieve the purpose of defense, by  the difference between the two in some specific subspaces. However, these detection methods are also vulnerable to adaptive attacks.

Moreover, a deficiency exists in the availability of a universally applicable and expeditious instrument for assessing the robustness of neural networks. A neural network is considered to be robust if it maintain stable and reliable performance in the face of changes in input data like adversarial examples. Many existing methods assess the robustness of neural networks based on the outcomes of adversarial attacks, a characterization deemed insufficiently accurate. Alternative methods striving for precision exhibit a considerable degree of complexity, necessitating a substantial amount of time for their implementation\cite{weng2018evaluating}\cite{singh2019abstract}. Pei et al.\cite{pei2017deepxplore} introduced a novel metric \textit{Neuron Coverage}(NC), which assesses network performance by the ratio of activated neurons to the total number of neurons for a given input. However, it has been demonstrated to be inaccurate\cite{harel2020neuron}.

In this work, we proposed a new metric called \textit{Neuron Coverage Change Rate}(NCCR), based on NC, to evaluate the robustness of a neural network. We observed that the reason for the imprecise evaluation of neural networks using NC is that inputs with different labels inherently possess distinct NC, while inputs with the same label exhibit similar NC. Therefore, NC can reflect the neural network's ability to distinguish inputs with different labels. We posit that a robust neural network should exhibit substantial disparities in NC for inputs with distinct labels, thereby increasing the difficulty for adversaries to generate adversarial examples. This presents an exploitable characteristic: considering an input $x$ and its counterpart $x^{'}$ obtained by introducing a small random perturbation $\epsilon$, a high robust neural network should generate similar NC for both, whereas a less robust network would exhibit the opposite behavior (even if the minor perturbation does not lead to incorrect output for $x^{'}$). The disparity in NC between these two inputs is termed NCCR. As we solely employed random perturbations in our methodology, the approach is expedient and independent of adversarial attacks. It effectively reflects the intrinsic properties of the neural network. 

As the computed NCCR is determined by the model, input, and the chosen perturbation epsilon, NCCR can also be employed to assess the robustness of the input when the epsilon and the model are fixed. One of the most commonly used methods for neural network detection of adversarial examples is to assess the robustness of input. Research indicates that adversarial examples, in order to maintain a sufficiently small difference from the original ones, tend to lie around the decision boundaries of the model. This results in significantly lower robustness of adversarial input compared to benign input\cite{zhao2021attack}. Furthermore, in the context of facing a backdoor attack, similar methods can be employed. A backdoor attack involves maliciously inserting a specific pattern or trigger condition into a neural network model, aiming to influence the model's behavior under specific conditions while maintaining normal performance in other scenarios, leading to deceptive outcomes. Similar to adversarial attack, the robustness of inputs with triggers in backdoor attacks is notably lower than that of benign inputs. This is because the majority of the contribution to the classification results comes from the trigger. Assuming the trigger is perturbed, it will lead to substantial changes in the output.

Our approach is not limited by task type: we conducted experiments in both image recognition and speaker recognition tasks, achieving commendable results in both domains. The experiments are divided into three parts: the first part focuses on using NCCR for robustness validation, with experiments conducted on MNIST\cite{lecun1998mnist} and CIFAR-10\cite{krizhevsky2009learning}; the second part involves adversarial examples detection, with experiments in image recognition tasks on MNIST\cite{lecun1998mnist}, CIFAR-10\cite{krizhevsky2009learning}, and ImageNet\cite{deng2009imagenet}, as well as speaker recognition tasks on Librispeech\cite{panayotov2015librispeech}; the third part addresses backdoor attack detection, with experiments conducted on GTSRB\cite{stallkamp2012man} and CIFAR-10\cite{krizhevsky2009learning}. In most of detection experiments, we achieved close to 100\% accuracy while requiring minimal computational resources. Numerous experiments demonstrate that our approach is not only more accurate but also more efficient. In a nutshell, we made the following contributions:
\begin{itemize}
\item We propose a novel metric called NCCR that enables rapid and accurate assessment of the robustness of neural networks.
\item We extend the application of NCCR to evaluate the robustness of model inputs, forming the basis for detecting adversarial and backdoor attacks.
\item Extensive experiments demonstrate that NCCR achieves higher accuracy in attack detection while requiring fewer computational resources.
\item We conducted experiments in both image recognition and speaker verification tasks, showcasing the versatility of NCCR for various downstream applications.
\end{itemize}

\section{Background}

\subsection{Robustness for DNN}

Taking an image recognition neural network as an example, consider an input image $x$ and an $L_{p}$-norm distance constraint $\epsilon$, allowing perturbations to be added to each pixel of the image to obtain $x^{'}$ while ensuring $||x^{'} - x||\leq \epsilon$. All such $x^{'}$ form a sphere centered at $x$ with a radius of $\epsilon$. If all images within this $L_{p}$-norm ball satisfy $f(x^{'} = f(x))$, meaning that any of the perturbed images within the ball does not alter the original classification result, we consider the neural network to be robust under the $L_{p}$-norm distance $\epsilon$. However, it is infeasible to exhaustively enumerate all possible images for testing. Therefore, various methods, including those establishing theoretical bounds or utilizing approximate solutions, have been proposed. These methods encompass complete verifiers such as SMT solving\cite{ehlers2017formal} and mixed integer linear programming\cite{tjeng2017evaluating}, as well as incomplete verifiers like abstract interpretation\cite{gehr2018ai2} and linear approximations\cite{weng2018towards}.

\subsection{Adversarial Attack}

If we can find any image in the $L_{p}$-norm ball within $\epsilon$ range that cause a change in the neural network's output, we have successfully achieved an \emph{adversarial attack}, and we refer to this input as an \emph{adversarial example}. \textbf{Fast Gradient Sign Method (FGSM)}\cite{goodfellow2014explaining} uses the derivative of the loss function $J(x, y)$ with respect to the input$x$ to increase the loss between $f(x)$ and the original label $y$: 
\[ x^{'} = x + \epsilon \times sign(\nabla_{x} J(x,y)) \]
where $\nabla$ is the partial derivative of the loss function$J(x,y)$ at $s$. Based on FGSM, Kurakin et al. proposed an iterative version called \textbf{Basic Iterative Method(BIM)}\cite{kurakin2018adversarial}. Instead of one-step approach like FGSM, BIM decomposes the attack into multiple steps, with each step computing the current optimal gradient and moving the input a smaller distance, reaching the $L_{p}$-norm threshold $\epsilon$ at last:

\begin{align*}
& x_{0} =x, \\
& x_{i + 1} = clip_{(x, \epsilon)}(x_{i} + \alpha\times sign(\nabla_{x_{i}}J(x_{i},y)))
\end{align*}
where $clip_{(x, \epsilon)}(\cdot)$ denotes the clipping function in order to constrain the perturbation $(x^{'} - x)$ within the $L_{p}$-norm threshold $\epsilon$. Generally, the gradient direction found by FGSM is not necessarily the optimal direction. Therefore, under the same $\epsilon$ constraint, BIM is more likely to find successful adversarial examples compared to FGSM.

Unlike the previous two gradient-based methods, Carlini and Wagner proposed the \textbf{C\&W}\cite{carlini2017towards} attack by transforming the adversarial attack problem into an optimization problem through the design of an objective function, for example:

\[
f(x^{'}) = max(max\{Z(x^{'}_{i}) : i \neq t\} - Z(x^{'}_{t}),-\kappa)
\]
where $Z(x^{'}_{i})$ is the softmax output of input $x$ for label $i$, $t$ is the target label and $\kappa$ is the hyperparameter used to balance the attack success rate(ASR) with the perturbation distance.

\subsection{Backdoor attack}

Different from adversarial attacks that do not require altering model parameters, a backdoor attack influences the generation of DNNs by contaminating the training set, allowing the DNN to exhibit predefined behaviors set by the attacker. During the training process, attacker injects a batch of data containing triggers into the training set, setting the labels of these instances uniformly to the desired one as specified by the attacker. This process results in the implantation of a backdoor within the DNN. The backdoor remains inconspicuous during normal operations of the DNN when presented with clean inputs. However, any input containing the designated trigger prompts the DNN to output the label predetermined by the attacker. In image recognition tasks, triggers may consist of imperceptible shapes formed by a small number of pixels or even semantically meaningful objects (e.g., glasses). The subtle nature of these features renders backdoors in DNNs challenging to detect.

\section{Neuron Coverage Change Rate}

\subsection{Characteristics of Neuron Coverage}
Our NCCR is based on the concept of Neuron Coverage proposed by Kexin Pei et al.\cite{pei2017deepxplore}: a threshold is artificially set, and when a example is input into the neural network, the ratio of neurons with outputs higher than this threshold to the total number of neurons is calculated. They argue that Neuron Coverage can reflect the completeness of testing for a network, where higher Neuron Coverage indicates a more comprehensive evaluation of the network. Their goal is to identify test cases that can detect erroneous behaviors in the network based on this criterion. However, we observed that neurons activated by inputs with the same label are very similar within the neural network, and there are distinct differences compared to inputs with different labels. This suggests that the neurons activated by inputs in the neural network should be relatively fixed, determined by their classification. Consequently, different classes of inputs are expected to have different Neuron Coverage. Therefore, blindly attempting to increase Neuron Coverage is unreasonable.

\begin{figure}[htbp]
  \centering

  \begin{subfigure}[b]{0.3\textwidth}
    \includegraphics[width=\textwidth]{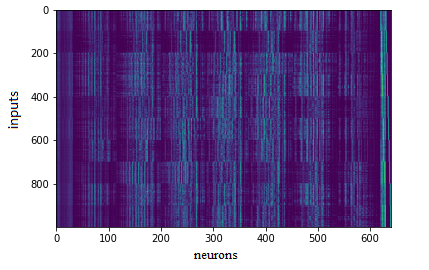}
    \caption{LeNet-1}
    \label{fig:subfig1}
  \end{subfigure}
  \hfill
  \begin{subfigure}[b]{0.3\textwidth}
    \includegraphics[width=\textwidth]{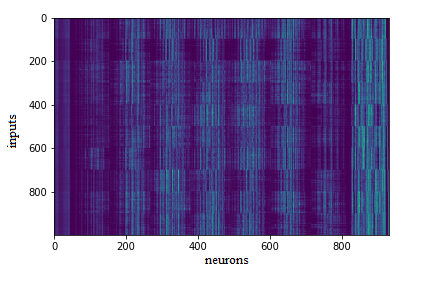}
    \caption{LeNet-4}
    \label{fig:subfig2}
  \end{subfigure}
  \hfill
  \begin{subfigure}[b]{0.3\textwidth}
    \includegraphics[width=\textwidth]{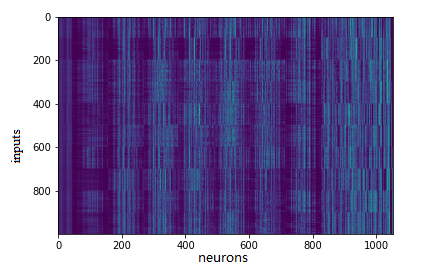}
    \caption{LeNet-5}
    \label{fig:subfig3}
  \end{subfigure}

  \caption{Neuron activation}
  \label{fig:ex1}
\end{figure}

We conducted experiments on three models trained on MNIST: Lenet-1, Lenet-4, and Lenet-5. For each class in the test set, we randomly selected 100 images, which is a total of 1000 images. The analysis involved recording the output produced by each neuron for every image. Figure \ref{fig:ex1} illustrates the results, with the x-axis denoting neurons in each network, the y-axis representing the 1000 selected images sorted by class, and the brightness of each point reflecting the activation value of the corresponding input in the respective neuron. Higher brightness is indicative of a higher activation value.

A noticeable stratification phenomenon is observed in all three figures, indicating that images with the same classification produce highly s imilar outputs on the same neuron within the neural network. Moreover, for all images, variations in the output of neurons can be observed, and situations where Neuron Coverage approaches 100\% are absent. 

\subsection{Definition of NCCR}

To address deficiencies in neuron coverage and achieve the goal of detecting erroneous behaviors in neural networks, we propose a new metric called NCCR. Unlike Neuron Coverage, which solely concerns itself with the intrinsic relationship between the network and the image, NCCR directs its attention towards the network's susceptibility to perturbations. \textbf{NCCR}: Given a trained neural network and an image \( x \) as input, we use \( F(x) \) to represent the set of outputs of each neuron after entering \( x \) into the neural network. Then, we add a random perturbation \( \epsilon \) to \( x \), resulting in \( x' \) and \( F(x') \). The degree of change in \( F(x') \) relative to \( F(x) \) is called NCCR. 
 
\section{Robustness Verification}
Since NCCR is directly related to the robustness of neural networks, we can utilize NCCR to assess the robustness of neural networks. A higher NCCR indicates a greater difference in neuron activations between \( x_0 \) and its perturbed version \( x_0' \), implying that the network is more sensitive to input changes and hence more vulnerable to adversarial attacks. Conversely, a lower NCCR suggests that the network maintains more stable neuron activations under perturbation, indicating stronger robustness. To evaluate the effectiveness of NCCR in measuring the robustness of neural networks, we conducted experiments on the MNIST and CIFAR-10 datasets. We first trained three models with varying levels of robustness on the MNIST and CIFAR-10 datasets: a standard model trained with natural data (denoted as \textbf{natural}), a model trained with adversarial examples generated by the Projected Gradient Descent (PGD) attack (\textbf{adv\_trained}), and a highly robust model trained using the Madry\cite{madry2017towards} method (\textbf{madry}). Different models with varying robustness levels all share the same structure: for MNIST, we use the LeNet-5 model, and for CIFAR-10, we use the ResNet model. We then evaluated the performance of these three models on three types of inputs: clean examples, adversarial examples generated through PGD attacks, and examples perturbed with random noise. The accuracy of each model on different datasets is shown in tables \ref{tab:accuracy on MNIST}and \ref{tab:accuracy on CIFAR10}. It is worth noting that adding random small perturbations to the clean examples only slightly affects the performance of the three models.

\begin{table}
\centering
\begin{tabular}{|l|l|l|l|}
\hline
                   & natural & adv\_trained & madry   \\\hline
clean examples     & 99.17\% & 98.40\%      & 98.53\% \\\hline
PDG attack         & 0\%     & 93.68\%      & 94.20\% \\\hline
randomly perturbed & 85.32\% & 93.75\%      & 97.70\% \\\hline
\end{tabular}
\caption{the accuracy of three models on MNIST}
\label{tab:accuracy on MNIST}
\end{table}

\begin{table}
\centering
\begin{tabular}{|l|l|l|l|}
\hline
                   & natural & adv\_trained & madry   \\\hline
clean examples     & 95.01\% & 87.25\%      & 87.14\% \\\hline
PDG attack         & 0\%     & 47.19\%      & 47.32\% \\\hline
randomly perturbed & 93.49\% & 87.24\%      & 87.08\% \\\hline
\end{tabular}
\caption{the accuracy of three models on CIFAR-10}
\label{tab:accuracy on CIFAR10}
\end{table}

\begin{figure}[h]
  \centering

  \begin{subfigure}[b]{0.48\textwidth}
    \includegraphics[width=\textwidth]{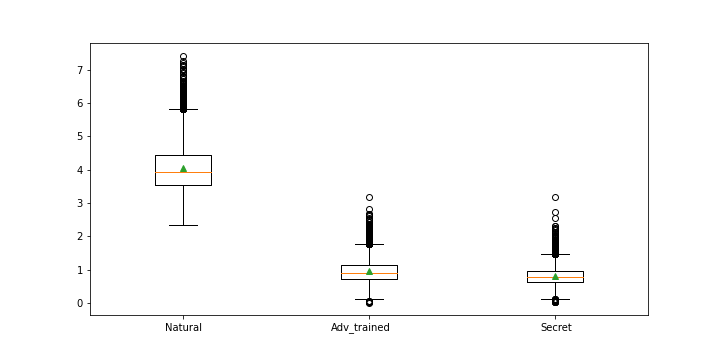}
    \caption{MNIST}
    \label{fig:MNIST}
  \end{subfigure}
  \hfill
  \begin{subfigure}[b]{0.48\textwidth}
    \includegraphics[width=\textwidth]{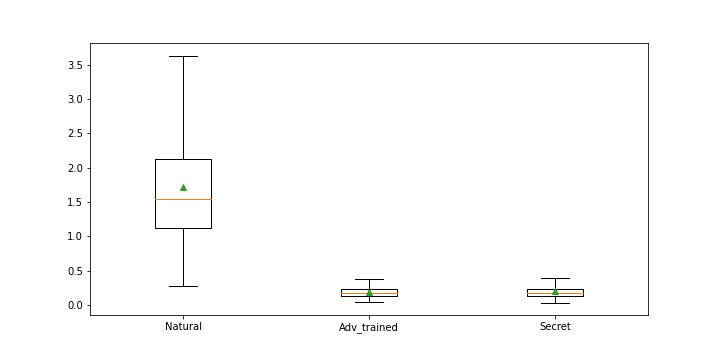}
    \caption{CIFAR-10}
    \label{fig:cifar10}
  \end{subfigure}

  \caption{NCCR of models with varying robustness}
  \label{fig:verification}
\end{figure}

To calculate the NCCR for each model, we generated ten sets of perturbed examples with small Gaussian noise. The perturbation ranges were \( |\epsilon| < 0.3/1 \) for MNIST and \( |\epsilon| < 8/255 \) for CIFAR-10. Both clean and perturbed examples were then input into each model, and the NCCR was calculated using the $l_2$-norm. The results are shown in the box plots in Figure~\ref{fig:verification}, where the yellow horizontal line represents the \textbf{median} and the green triangle indicates the \textbf{mean} of the NCCR values. The whiskers display the range, and the circles represent \textbf{outliers}. From the box plots, we observe that the low-robustness \textbf{natural} model consistently produces significantly higher NCCR values compared to the two more robust models, \textbf{adv\_trained} and \textbf{madry}. Additionally, \textbf{madry} model shows slightly lower NCCR values than \textbf{adv\_trained} model, with this difference being more pronounced in the MNIST dataset. These findings indicate that NCCR effectively reflects the robustness of the models, with the natural model, being less robust, exhibiting a higher NCCR, while the adversarially trained models demonstrate lower NCCR values, highlighting their increased resistance to perturbations. Furthermore, since the NCCR is calculated using the test set from the trained models along with non-adversarial perturbed examples, which are created by adding small Gaussian noise, this method serves as a model robustness evaluation that is independent of the attack strategy.

\section{Adversarial Examples Detection}

In the results of the previous section, we also observed an interesting phenomenon: for the same model, there can be significant variation in the NCCR values across different images. Even for the low-robustness \textbf{natural} model, some images can produce very low NCCR values, which may even match the general NCCR values of the other two high-robustness models. This is because the NCCR generated by a given input reflects not only the robustness of the model but also the inherent robustness of the input itself. For a particular image, its robustness refers to the ease with which its label changes when subjected to attacks. According to Zhao et al. \cite{zhao2021attack}, adversarial examples typically exhibit lower robustness than clean examples because they are usually located near the model's decision boundary. This means that, while being adversarial, they are also highly susceptible to further attacks; small perturbations can push these adversarial examples across the decision boundary and change their label.

Based on this property, we can use NCCR for adversarial example detection. Given a trained model for adversarial example detection, we compute a baseline NCCR using a test set \(X_0\), which represents the NCCR for clean examples. This means that an input example without any attack should have an NCCR similar to \(NCCR_0\). If we obtain an NCCR significantly higher than the baseline, it indicates that the input example has much lower robustness compared to clean examples, and therefore, it is likely to be an adversarial example.

\subsection{Detection For Image Recognition Models}

In this section, we will implement adversarial example detection for image recognition networks using NCCR. For an image recognition network, we first generate a set of adversarial examples based on a clean dataset using a widely used attack method. Then, we calculate the NCCR for both the clean examples and the adversarial examples, and use these NCCR values to train a classifier. Once the classifier is trained, it can classify an input as either a clean example or an adversarial example based on the input's NCCR.

\begin{figure}
    \centering
    \includegraphics[width=0.6\linewidth]{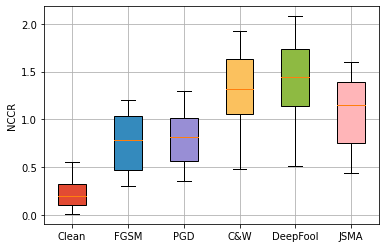}
    \caption{Comparison of NCCR between Clean Examples and Adversarial Examples}
    \label{fig:nccr_compare}
\end{figure}

However, merely calculating the $l_2$ or $l_0$ distance of the NCCR of the examples is not sufficient to completely distinguish between adversarial examples and clean examples. Taking the CIFAR-10 dataset as an example, the $l_2$ distance NCCR of each image and its corresponding adversarial example is listed as a whole in \ref{fig:nccr_compare}. It can be seen that although the overall NCCR of clean examples is smaller than that of adversarial examples, there is no complete separation: the largest NCCR among all clean examples is still greater than the smallest NCCR among all adversarial examples. This indicates that directly calculating the NCCR to approximate the upper and lower bounds of example robustness is overly broad. The error caused by the loss of information in the distance calculation makes the NCCR an unsuitable indicator for modeling robustness and detecting adversarial examples.

When training the NCCR classifier, pre-generated adversarial examples are required. Regarding the selection of attack methods for generating adversarial examples used in training, this paper hopes that the chosen attack can represent the category with the highest robustness among all attack methods. If the trained classifier can demonstrate excellent discrimination ability against adversarial examples with high robustness, it will also work well against other adversarial examples with lower robustness. Based on this consideration, according to Figure \ref{fig:nccr_compare}, this paper selects FGSM, PGD, JSMA, and Combine, which includes all three types of adversarial examples, as the adversarial attacks used for training the classifier.

Specifically, we have a trained image recognition network \( F(x) \) and a dataset \( X_0 \) used during training. First, we use an attack method \( \alpha \) to generate adversarial examples \( X_0' \) based on \( X_0 \). We then compute the NCCR for both \( X_0 \) and \( X_0' \). These NCCR values, \( \text{NCCR}(X_0) \) and \( \text{NCCR}(X_0') \), are used as inputs to train a classifier \( D \). The classifier \( D \) learns to distinguish between the different distributions of NCCR values for clean and adversarial examples, allowing it to classify whether a given NCCR corresponds to a clean example or an adversarial one.

During the use of \( F(x) \), we first calculate the NCCR for each input \( x \) and use the classifier \( D \) to determine whether it is a clean example or an adversarial example. If it is a clean example, the network outputs the normal class of \( x \); if it is an adversarial example, a warning is issued, thereby providing a defense mechanism for the network.

\subsection*{Experiment Setups}

We conducted experiments on MNIST, CIFAR-10, and ImageNet to validate the feasibility of our method and compare it with three baseline approaches. 

\textbf{Baseline 1} \cite{wang2019adversarial}proposes detecting adversarial examples by evaluating the sensitivity of input examples to random mutations of the deep neural network (DNN) model. Sensitivity is measured using the \textit{Label Change Rate} (LCR), which quantifies the degree of change in the classification result after the mutation of the input. Adversarial examples generally exhibit a higher LCR compared to normal examples.

\textbf{Baseline 2} \cite{wang2020dissector}works by monitoring the input verification process within the deep learning model. It employs a fault-tolerant approach to distinguish between "within-inputs" (normal inputs that the model can handle) and "beyond-inputs" (inputs the model cannot process). This method ensures that the confidence of the model's predictions increases steadily for normal inputs, while adversarial inputs disrupt this confidence progression.

\textbf{Baseline 3} \cite{zhao2021attack}detects adversarial examples based on differences in example robustness. This method evaluates how difficult it is to attack an input example, with adversarial examples being more easily attacked due to their lower robustness.

The first experimental environment $Env\_1$ is provided by $BL_1$, which includes a LeNet-5 model based on MNIST and a GoogLeNet model based on CIFAR-10. Their accuracy rates on the test set are 98.3\% and 90.5\%, respectively. The second experimental environment $Env_2$ is provided by $BL_2$, which includes a LeNet-4 model based on MNIST, a WRN-28-10 model based on CIFAR-10, and a ResNet-101 model based on ImageNet. Their accuracy rates (ImageNet uses top-1 accuracy) on the test set are 98.4\%, 96.2\%, and 77.36\%, respectively. $BL_3$ has been implemented in both environments, so it is compared with $BL_1$ and $BL_3$ in $Env_1$, and with $BL_2$ and $BL_3$ in $Env_2$. To visually demonstrate the detection performance of different methods, this paper uses AUROC as the detection metric.

We used Foolbox\cite{rauber2017foolbox} to generate adversarial examples, with all parameters set to their default values.

The classifier is a simple three-layer fully connected network with a total of 1792 parameters. The input to the classifier is a vector representing the NCCR of the image \( x \), which is calculated using the model \( F(x) \). The output of the classifier is a binary decision, indicating whether \( x \) is an adversarial example or a clean example.

\subsection*{Evaluation}
Table\ref{table:env1} and table \ref{table:env2} presents the classification performance of classifiers trained with different attack methods on various adversarial examples, along with a comparison to the three baseline methods. We use the AUROC (Area Under the Receiver Operating Characteristic Curve) to measure classification performance. A higher AUROC indicates that the classifier is better at distinguishing between clean and adversarial examples. Each column of the table represents an attack method used to train the classifier, namely the previously mentioned FGSM, PGD, JSMA, and a combined version of the three attack methods. The baseline row indicates the best result from the three baseline methods for each corresponding group.

Overall, our method outperforms the best baseline in most cases. In the MNIST experiment, the results are all very close to 1.0, so our advantage is not as apparent. However, in the CIFAR-10 and ImageNet experiments, our method significantly outperforms the baseline, particularly in the combined attack group, where we achieve better results than the baseline in most cases.

It is worth noting that when detecting adversarial examples generated by the JSMA attack, the performance of our method is generally modest. We only achieved good results when detecting adversarial examples using classifiers trained with the same JSMA attack (which, unsurprisingly, yields better results). This is because JSMA is an $l_0$ attack, meaning that only a small number of pixels are heavily modified in the generated adversarial example. In contrast, the perturbations we use to compute the NCCR are random noise, an $l_2$ attack method. As a result, our method performs slightly worse when detecting $L_0$ attacks compared to other attack types.

\begin{table}[]
\centering
\begin{tabular}{c|l|l|l|l|l|l|l}
\hline
\multicolumn{1}{l|}{$Env_1$} & attack   & FGSM            & PGD             & JSMA   & Combine & $BL^{IMG}_1$    & $BL^{IMG}_3$             \\ \hline
\multirow{5}{*}{MNIST}    & FGSM     & 0.9943          & \textbf{0.9949} & 0.9229 & 0.9935  & 0.9617 & 0.9883          \\ \cline{2-8} 
                          & PGD      & 0.9926          & \textbf{0.9973} & 0.9131 & 0.9926  & 0.9531 & 0.9783          \\ \cline{2-8} 
                          & JSMA     & 0.9217          & 0.9171          & 0.9537 & 0.9164  & 0.9941 & \textbf{0.9984} \\ \cline{2-8} 
                          & C\&W     & \textbf{0.9950} & 0.9932          & 0.9558 & 0.9807  & 0.9576 & 0.9870          \\ \cline{2-8} 
                          & DeepFool & \textbf{0.9983} & 0.9964          & 0.9672 & 0.9851  & 0.9817 & 0.9971          \\ \hline
\multirow{5}{*}{CIAR-10}  & FGSM     & \textbf{0.9974} & 0.9951          & 0.9643 & 0.9899  & 0.8617 & 0.8998          \\ \cline{2-8} 
                          & PGD      & 0.9906          & \textbf{0.9916} & 0.9612 & 0.9891  & 0.8741 & 0.9014          \\ \cline{2-8} 
                          & JSMA     & 0.8931          & 0.9002          & 0.9415 & 0.9103  & 0.9682 & \textbf{0.9890} \\ \cline{2-8} 
                          & C\&W     & \textbf{0.9989} & 0.9987          & 0.9843 & 0.9964  & 0.9063 & 0.9176          \\ \cline{2-8} 
                          & DeepFool & \textbf{0.9971} & 0.9883          & 0.9801 & 0.9904  & 0.9614 & 0.9902          \\ \hline
\end{tabular}
\caption{AUROC Comparison with the Baseline Method in $Env_1$}
\label{table:env1}
\end{table}

\begin{table}[]
\centering
\begin{tabular}{c|l|l|l|l|l|l|l}
\hline
\multicolumn{1}{l|}{$Env_2$} & attack   & FGSM            & PGD             & JSMA            & Combine & $BL^{IMG}_2$    & $BL^{IMG}_3$             \\ \hline
\multirow{5}{*}{MNIST}    & FGSM & \textbf{1.0}    & \textbf{1.0} & 0.9981          & 0.9999 & 0.9993 & \textbf{1.0} \\ \cline{2-8} 
                          & PGD      & 0.9997 & \textbf{1.0}    & 0.9979          & 0.9998  & 0.9992 & \textbf{1.0}    \\ \cline{2-8} 
                          & JSMA     & 0.9118          & 0.9226          & 0.9465          & 0.9212  & 0.9993 & \textbf{0.9999}          \\ \cline{2-8} 
                          & C\&W     & \textbf{1.0}    & \textbf{1.0}    & 0.9878          & 0.9935  & 0.9996 & \textbf{1.0}    \\ \cline{2-8} 
                          & DeepFool & \textbf{0.9982} & 0.9976 & 0.9751          & 0.9924  & 0.9892 & 0.9877          \\ \hline
\multirow{5}{*}{CIAR-10}  & FGSM & \textbf{0.9995} & 0.9945       & \textbf{0.9995} & 0.9991 & 0.9981 & 0.9983       \\ \cline{2-8} 
                          & PGD      & 0.9993          & \textbf{0.9996} & 0.9943          & 0.9987  & 0.9974 & 0.9972          \\ \cline{2-8} 
                          & JSMA     & 0.9054          & 0.9841          & \textbf{0.9972} & 0.9802  & 0.9966 & 0.9962          \\ \cline{2-8} 
                          & C\&W     & 0.9526          & 0.9975          & 0.9948          & 0.9899  & 0.9968 & \textbf{0.9985} \\ \cline{2-8} 
                          & DeepFool & 0.9857          & \textbf{0.9808} & 0.9803          & 0.9786  & 0.9618 & 0.9713          \\ \hline
\multirow{5}{*}{ImageNet} & FGSM & \textbf{0.9813} & 0.9805       & 0.9714          & 0.9788 & 0.9617 & 0.9782       \\ \cline{2-8} 
                          & PGD      & 0.9820          & \textbf{0.9825} & 0.9641          & 0.9766  & 0.9562 & 0.9696          \\ \cline{2-8} 
                          & JSMA     & 0.9341          & 0.9223          & 0.9637          & 0.9415  & 0.9695 & \textbf{0.9962} \\ \cline{2-8} 
                          & C\&W     & \textbf{0.9974} & 0.9953          & 0.9744          & 0.9904  & 0.9636 & 0.9924          \\ \cline{2-8} 
                          & DeepFool & \textbf{0.9961} & 0.9947          & 0.9710          & 0.9917  & 0.9924 & 0.9958 \\ \hline
\end{tabular}
\caption{AUROC Comparison with the Baseline Method in $Env_2$}
\label{table:env2}
\end{table}

\subsection{Detection For Speaker Recognition Models}

In addition, we have implemented adversarial examples detection using NCCR in the field of speaker recognition. Speaker recognition refers to the process of verifying a person's identity by analyzing their vocal features. Unlike traditional biometric techniques such as passwords and fingerprints, speaker recognition utilizes physiological and behavioral traits in speech, such as vocal cord vibration patterns, intonation, speech rate, and pronunciation habits, to authenticate an individual's identity. Its applications are widespread, covering areas such as security authentication, intelligent assistants, and speech monitoring.

Despite some similarities in technical implementation between speaker recognition and image recognition, such as both relying on deep learning models for feature extraction and classification, significant differences exist between the two. First, speaker recognition primarily deals with continuous time-series data (speech signals), whereas image recognition processes static two-dimensional pixel data. Second, speech data is often influenced by noise, environmental changes, and emotional fluctuations, which makes the challenges in data preprocessing and feature extraction more complex in speaker recognition compared to image recognition. Lastly, adversarial attacks in image recognition generally rely on pixel-level perturbations, while adversarial attacks in speaker recognition may manipulate features such as the speech spectrum, pitch, and pronunciation to deceive the system, making these attacks more subtle and intricate.

In our experiments on speaker recognition, the methodology we employed was largely similar to that in image recognition: we used pre-generated adversarial examples and clean examples to compute the NCCR, which was then used to train a classifier. The trained classifier was able to distinguish between clean examples and adversarial examples.

\subsection*{Experiment Setups}

\textbf{Dataset:} We selected 251 individuals from the LibriSpeech dataset\cite{panayotov2015librispeech} for training and testing. The test set contains a total of 25,652 speech examples, and among these, 2,887 examples are used for evaluation.

\textbf{Model:} We used two speaker recognition networks:
1. The first model is the \textbf{xVector-PLDA} \cite{dehak2010front} network. This model is an improved version of the traditional iVector, which provides higher performance. In contrast to iVector, which typically uses GMM to extract speaker-related features, xVector utilizes TDNN for processing. These vectors are generally more discriminative than iVector and better capture the unique characteristics of a speaker.
2. The second model is a one-dimensional Convolutional Neural Network \textbf{AudioNet} proposed in \cite{becker2024audiomnist}. This model's structure is more similar to the one used in image recognition tasks.

\textbf{Attack Methods:} Adversarial examples were generated using the \texttt{SpeakerGuard} platform developed by Chen et al. The attack methods used include PGD, CW2, and FAKEBOB.

\textbf{Baseline:} We compared our results with two existing works. In \cite{wu2022adversarial}, the method first extracts features from the audio signal (e.g., Mel spectrograms), then synthesizes the features back into audio using a trained vocoder. By comparing the ASV (Automatic Speaker Verification) score difference between the original and synthesized audio, the method effectively distinguishes between real and adversarial examples. For real examples, the score difference is small, while for adversarial examples, the score difference is significant. In \cite{chen2023detection}, the MEH-FEST detection method is proposed. The detection principle relies on three main observations: 
- Adversarial perturbations broadly affect the audio signal like white noise, especially during periods of silence.
- Audio signals are non-stationary, with significant differences in the impact of perturbations when speech is present versus when it is absent.
- The energy of original audio is typically low in high-frequency bands, especially when there is no speech present.

\subsection*{Evaluation}

The experimental results are shown in the figure. Overall, our method outperforms the baseline in most attack scenarios. Specifically, for CW and PGD attacks, the results are comparable to those of the vocoder-based method, but for FAKEBOB, our method clearly outperforms the vocoder. On the other hand, MEH-FEST performs well in detecting FAKEBOB, but shows poor performance in detecting the other two attack types. Therefore, our method demonstrates better generalizability.

\begin{table}[]
\centering
\begin{tabular}{c|l|l|l|l|l}
\hline
\multicolumn{1}{l|}{env}   & attack  & FGSM            & PGD             & $BL^{ASV}_1$    & $BL^{ASV}_2$ \\ \hline
\multirow{4}{*}{$Env_e$}    & FGSM    & 0.9621          & \textbf{0.9644} & 0.7648          & 0.9164       \\ \cline{2-6} 
                            & PGD     & \textbf{0.9659} & 0.9427          & 0.7513          & 0.9026       \\ \cline{2-6} 
                            & C\&W    & \textbf{0.9714} & 0.9694          & 0.8647          & 0.9457       \\ \cline{2-6} 
                            & FAKEBOB & 0.9613          & 0.9537          & \textbf{0.9861} & 0.9014       \\ \hline
\multirow{4}{*}{$Env_{ne}$} & FGSM    & 0.9671          & \textbf{0.9716} & 0.7134          & 0.8988       \\ \cline{2-6} 
                            & PGD     & 0.9646          & \textbf{0.9968} & 0.7652          & 0.8859       \\ \cline{2-6} 
                            & C\&W    & \textbf{0.9785} & 0.9693          & 0.6945          & 0.9145       \\ \cline{2-6} 
                            & FAKEBOB & 0.9581          & 0.9474          & \textbf{0.9816} & 0.8416       \\ \hline
\end{tabular}
\caption{Comparison of Adversarial Detection Methods for ASV}
\label{tabel:ASV_Detection}
\end{table}

\section{Backdoor Attack Detection} 

Besides the adversarial attack detection mentioned above, NCCR can also be applied to backdoor attack detection. In this section, we will introduce the implementation of backdoor attack detection using NCCR to distinguish the robustness difference between examples with and without trigger.

\subsection{Backdoor Detection Analysis}

A backdoor attack involves embedding a trigger or malicious pattern into a machine learning model during its training phase. Unlike adversarial attacks, where the goal is to manipulate the model's behavior in response to specific inputs, backdoor attacks aim to cause the model to misbehave only when a specific "trigger" is present in the input. The backdoor is "hidden" during normal operation and is activated only when the trigger appears, allowing the attacker to control the model's predictions for particular inputs. Although the principles behind trigger-based attacks are not exactly the same, we can still approach the problem from the perspective of image robustness. The key to a backdoor attack lies in determining whether an image contains a trigger. If an image does not contain a trigger, the network will classify the image according to its typical classification process. However, once the network detects the presence of a trigger in the image, the input image will be classified with high confidence into the attacker’s pre-defined class.

To ensure the attack’s stealthiness, the attacker typically sets the trigger to be very small, making it difficult to notice. This also means that the trigger is easily disrupted, and a small perturbation can cause a significant change in the confidence level of the image’s original class, which is much larger than for clean examples. Therefore, images containing a trigger will exhibit significantly lower robustness when faced with a perturbation large enough to destroy the trigger, and this feature is similar to that of adversarial examples: adversarial examples have significantly lower robustness than clean examples. Based on this insight, we can apply a similar approach to detect backdoor attacks.

However, different from the adversarial examples, the decision space of the poisoned examples in the backdoor attack is significantly different from that of the clean examples. In the adversarial attack, the adversarial example takes a short distance in the decision space and crosses the decision boundary to change the prediction result after a small perturbation of the clean example. The adversarial example is usually low robustness because it is too close to the decision boundary. However, in backdoor attacks, poisoned examples can be very far from the decision boundary, because the model usually has a high confidence in the trigger, which means that these poisoned examples are significantly more robust than clean examples like BadNets\cite{gu2019badnets}); In some backdoor attacks, especially those that pursue high concealment, the triggers are often designed to be small and fragile in order to avoid inspection, such as Blended attack\cite{chen2017targeted} with low weight triggers. This kind of attack will also be less robust than clean examples due to the easy destruction of triggers.

In order to solve the problem of inconsistent robustness of the poisoned examples of different backdoor attacks, this chapter proposes to use a sufficiently strong perturbation in the calculation of NCCR such that all triggers will be destroyed. In this case, even highly robust explicit trigger examples such as BadNets produce significantly higher NCCR than normal examples.

\begin{figure}[]
    \centering
    \begin{subfigure}[b]{0.45\textwidth}
        \includegraphics[width=\textwidth]{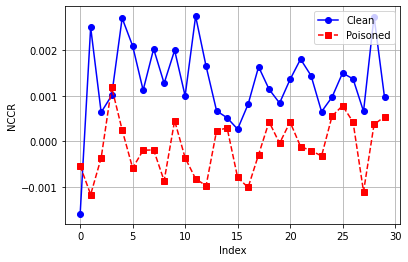}
        \caption{NCCR Comparison with $\epsilon$ = 8}
        \label{fig:NCCR_linear_wave_high}
    \end{subfigure}
    \hfill
    \begin{subfigure}[b]{0.45\textwidth}
        \includegraphics[width=\textwidth]{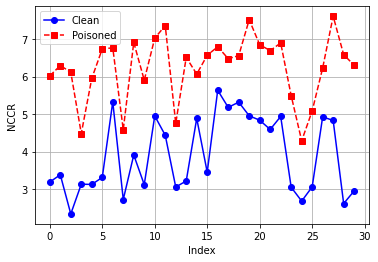}
        \caption{NCCR Comparison with $\epsilon$ = 256}
        \label{fig:NCCR_linear_wave_low}
    \end{subfigure}
    \caption{NCCR Comparison of the Poisoned and Clean Exexamples under Different Perturbations}
    \label{fig:NCCR_linear_wave}
\end{figure}

Here is an example, the VGG11 model is used to train a model with a backdoor, the attack method is BadNets, where the trigger is the white square of 5$\times$5 in the lower right corner of the image, and the attack target is 33 labels. After training, the accuracy of the model in the clean test set is 97.35\%, and the accuracy in the poisoned test set is 98.09\%. Select the first 1,000 images in the test set and add a trigger to generate the corresponding poisoned examples. Calculate and average the NCCR of the clean and poisoned examples when $\epsilon= 8$(small perturbation) and $\epsilon=256$(complete destruction of image structure), respectively. We get 1,000 NCCR (scalar) for each of the clean and infected examples. Figure \ref{fig:NCCR_linear_wave} shows the average NCCR values of clean examples and infected examples for the first 30 groups of images when $\epsilon$is 8 and 256, respectively, which intuitively reflects the inversion of NCCR difference. From \ref{fig:NCCR_linear_wave_low} it is clear that the NCCR of the poisoned example is lower than that of the clean example when $\epsilon$is very small, but the NCCR order of magnitude is $10^{-3}$, so the difference is not large. When $\epsilon$is large enough to completely destroy the image content, the \ref{fig:NCCR_linear_wave_high} shows that the NCCR of the poisoned example is significantly higher than that of the clean example, even though both image contents are  no longer visually recognizable.

\begin{figure}
    \centering
    \includegraphics[width=0.6\linewidth]{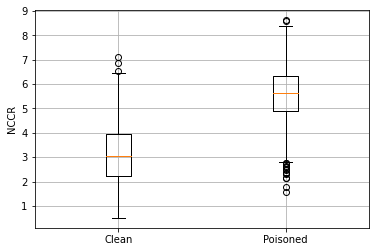}
    \caption{Overall NCCR Comparison when $\epsilon$=256}
    \label{fig:NCCR_box}
\end{figure}

The mean value of NCCR alone is not enough to completely distinguish the infected examples from the clean examples. Figure \ref{fig:NCCR_box} gives a visual representation of the overall distribution of the mean NCCR values between the clean and the infected examples when $\epsilon$=256.

However, in the detection method of backdoor attacks, we cannot use the pre-generated adversarial examples to calculate the NCCR training classifier, and use the binary classification decoder to automatically identify it as in the adversarial example detection method. In practical application scenarios, it is impossible for the model training party to know whether the model contains backdoors and trigger styles in advance, so the corresponding toxic examples cannot be generated in advance. In order to distinguish between the NCCR of the clean examples and the NCCR of the poisoned examples, we will use a clustering method to try to separate the NCCR of all the training sets into two clusters, and then evaluate whether the clusters are reasonable using the silhouette coefficient.

\subsection{Experiment Setups}

\textbf{Dataset:} We conducted experiments on the CIFAR-10 and GTSRB datasets.

\textbf{Model:} 
We use ResNet-34 for both datasets.

\textbf{Poisoning Method:} We implemented backdoor insertion using the BadNets, Blended, SIG, ReFool based on the \texttt{BackdoorBox} framework, to implant backdoors into the models.

\textbf{Baseline:} We compared our results with two baseline methods:
1. The method proposed in \cite{gao2019strip}, called STRIP, detects backdoors by observing the entropy of the predicted class after perturbation. Specifically, when a example is attacked with a Trojan backdoor, the perturbed input's prediction exhibits low entropy, while for normal inputs, the perturbed predictions typically show higher entropy.
2. The method proposed in \cite{chen2018detecting} is called Activation Clustering. This method detects backdoors by reducing the dimensionality and analyzing the activations of the neural network while processing training data, in order to identify whether the data has been maliciously modified by an inserted backdoor trigger.

\subsection*{Evaluation}

To evaluate the detection effectiveness of different methods, this chapter uses the F1 score as the evaluation metric. 

The experimental results are shown in Table \ref{table:backdoor_results}. In all the experiments, the NCCR detection method achieved the best results. The detection results of the activation clustering method were very close to those of NCCR, but never exceeded those of the NCCR method. The STRIP method had good detection performance for fixed-form triggers, but its detection performance for more variable and invisible attacks was much worse than that of the NCCR detection method. The NCCR detection method achieved an F1-score of nearly 1 in the detection results for each type of attack, which means the accuracy was almost 100\%. This indicates that the NCCR detection method has strong generalization ability and can achieve precise detection for different attack methods.

\begin{table}[]
\centering
\begin{tabular}{c|l|l|l|l}
\hline
\multicolumn{1}{l|}{datasets}  & attacks  & NCCR            & AC & STRIP  \\ \hline
\multirow{4}{*}{CIFAR-10} & BadNets & \textbf{0.9994} & 0.9915      & 0.9884 \\ \cline{2-5} 
                          & Blended & \textbf{0.9995} & 0.9874      & 0.7416 \\ \cline{2-5} 
                          & SIG     & \textbf{0.9989} & 0.9867      & 0.6451 \\ \cline{2-5} 
                          & ReFool  & \textbf{0.9986} & 0.9833      & 0.6846 \\ \hline
\multirow{4}{*}{GTSRB}    & BadNets & \textbf{0.9998} & 0.9991      & 0.9798 \\ \cline{2-5} 
                          & Blended & \textbf{0.9996} & 0.9982      & 0.7064 \\ \cline{2-5} 
                          & SIG     & \textbf{0.9994} & 0.9978      & 0.6291 \\ \cline{2-5} 
                          & ReFool  & \textbf{0.9993} & 0.9986      & 0.6487 \\ \hline
\end{tabular}
\caption{Backdoor attack detection results}
\label{table:backdoor_results}
\end{table}

\section{Conclusion}

We propose a metric, NCCR, to measure the robustness of both the network and input examples. First, we use NCCR to validate the robustness of different models under identical inputs. Next, we apply NCCR to detect adversarial examples in both image recognition and speaker recognition models by evaluating the robustness of the input examples. Finally, we extend the use of NCCR to detect triggers for backdoor attacks.

\section{To Do}

1. Show the experimental results in more detail; 2. Time cost of different methods; 3. Discussion on the effect difference of two models in Speaker recognition

\bibliography{reference}

\end{document}